# Destabilizing turbulence in pipe flow

Jakob Kühnen[1*]†, Baofang Song[1,2*], Davide Scarselli[1], Nazmi Burak Budanur[1], Michael Riedl[1], Ashley Willis[3], Marc Avila[2] and Björn Hof[1]†

[1]Nonlinear Dynamics and Turbulence Group, IST Austria, 3400 Klosterneuburg, Austria

[2]Center of Space Technology and Microgravity (ZARM), Universität Bremen, 28359 Bremen, Germany

[3]School of Mathematics and Statistics, University of Sheffield, S3 7RH, United Kingdom

* These authors contributed equally to this work

†To whom correspondence should be addressed: jakob.kuehnen@ist.ac.at, bhof@ist.ac.at

**Turbulence is the major cause of friction losses in transport processes and it is responsible for a drastic drag increase in flows over bounding surfaces. While much effort is invested into developing ways to control and reduce turbulence intensities (*1–3*), so far no methods exist to altogether eliminate turbulence if velocities are sufficiently large. We demonstrate for pipe flow that appropriate distortions to the velocity profile lead to a complete collapse of turbulence and subsequently friction losses are reduced by as much as 95%. Counterintuitively, the return to laminar motion is accomplished by initially increasing turbulence intensities or by transiently amplifying wall shear. The usual measures of turbulence levels, such as the Reynolds number (Re) or shear stresses, do not account for the subsequent relaminarization. Instead an amplification mechanism (*4*, *5*) measuring the interaction between eddies and the mean shear is found to set a threshold below which turbulence is suppressed beyond recovery.**

Flows through pipes and hydraulic networks are generally turbulent and the friction losses encountered in these flows are responsible for approximately 10% of the global electric energy consumption. Here turbulence causes a severe drag increase and consequently much larger forces are needed to maintain desired flow rates. In pipes, both laminar and turbulent states are stable (the former is believed to be linearly stable for all Re, the latter is stable if Re>2040 (*6*)), but with increasing speed the laminar state becomes more and more susceptible to small disturbances. Hence in practice most flows are turbulent at sufficiently large Re. While the stability of laminar flow has been studied in great detail, little attention has been paid to the susceptibility of turbulence, the general assumption being that once turbulence is established it is stable.

Many turbulence control strategies have been put forward to reduce the drag encountered in shear flows (*7–17*). Recent strategies employ feedback mechanisms to actively counter selected velocity components or vortices. Such methods usually require knowledge of the full turbulent velocity field. In computer simulations (*7*),(*8*) it could be demonstrated that under these ideal conditions, flows at low Reynolds number can even be relaminarized. In experiments the required detailed manipulation of the time dependent velocity field is, however, currently impossible to achieve. Other studies employ passive (e.g. riblets) or active (oscillations or excitation of travelling waves) methods to interfere with the near wall turbulence creation. Typically here drag reduction of 10 to 40 percent has been reported, but often the control cost is substantially higher than the gain, or a net gain can only be achieved in a narrow Reynolds number regime.

Instead of attempting to control or counter certain components of the complex fluctuating flow fields, we will show in the following that by appropriately disturbing the mean profile, turbulence can be pushed outside its limit of existence and as a consequence the entire flow relaminarizes. Disturbance schemes are developed with the aid of direct numerical simulations (DNS) of pipe flow and subsequently implemented and tested in experiments. In the DNS a flow is simulated in a five diameter (D) long pipe and periodic boundary conditions are applied in the axial direction. Initially we perturb laminar pipe flow by adding fluctuation levels of a fully turbulent velocity field rescaled by a factor (f) to a laminar flow field. As shown in Fig.1a (dark blue curve), for small initial perturbations, i.e. small f, the disturbance eventually decays and the flow remains laminar. For sufficiently large amplitudes (f of order unity) turbulence is triggered (purple, red and cyan curves). So far this is the familiar picture of the transition to turbulence in shear flows, where turbulence is only triggered if perturbation amplitudes surpass a certain threshold. However, when increasing the turbulent fluctuations well beyond their usual levels (f>2.5), surprisingly the highly turbulent flow almost immediately collapses and returns to laminar (light and dark green curves). Here the initially strong vortical motion leads to a redistribution of shear resulting in an unusually flat velocity profile (black profile in Fig.1c).

To achieve a similar effect in experiments we increase the turbulence level by vigorously stirring a fully turbulent pipe flow (Re=3500), employing four rotors located inside the pipe 50 D downstream of the pipe inlet (Extended Data Fig. 1). As the highly turbulent flow proceeds further downstream it surprisingly does not return to the normal turbulence level but instead it quickly reduces in intensity until the entire flow is laminar (Fig.1 b top to bottom and supplementary movie). Being linearly stable the laminar flow persists for the entire downstream pipe. In a second experiment, turbulent flow (Re=3100) is disturbed by injecting fluid through 25 small holes (0.5mm diameter) in the pipe wall (holes are distributed across a pipe segment with a length of 25D, (see Methods and Extended Data Fig.3)). Each injected jet creates a pair of counter-rotating vortices, intensifying the eddying motion beyond the levels of ordinary turbulence at this Re. The additional vortices redistribute the flow and as a consequence the velocity profile is flattened (Fig. 1c, purple dotted line). When the perturbation is actuated downstream fluctuation levels drop and the centre line velocity returns to its laminar value (Fig.1 d). Laminar motion persists for the remainder of the pipe. In this case the frictional drag is reduced by a factor of 2. Overall the injected fluid only amounts to ~1.5 % of the total flow rate in the pipe. Likewise the kinetic energy required for relaminarization (including the actuation and the drag increase in the control section) is only ~2.6% of the energy saved because of the drag decrease in the remainder of the pipe and the net power saving amounts to 45% (see Methods).

In another experiment we attempted to disrupt turbulence (Re=5000) by injecting fluid parallel to the wall in the streamwise direction (see Methods and Extended Data Fig. 2 and 4). Unlike for the previous case, this disturbance does not result in a magnification of cross-stream fluctuations, but instead increases the wall shear stress and hence also the friction Reynolds number, $Re_\tau$. Directly downstream of the injection point the latter is increased by about 15%. The acceleration of the near wall flow automatically causes deceleration of the flow in the pipe centre (the overall mass flux is held constant) hence again resulting in a flatter velocity profile (blue in Fig.1c). Despite the local increase in $Re_\tau$, further downstream the fluctuation levels begin to drop and the turbulent flow has been sufficiently destabilized that eventually

(30 D downstream) it decays and the flow returns to laminar. As a result, friction losses drop by a factor of 2.9 (see Fig.2a) and a net power saving of 55 % is achieved (see Methods). For this type of perturbation we find, that relaminarization occurs for an intermediate injection range (~15% of the flow rate in the pipe), while for smaller and larger rates the flow remains turbulent. A property common to all above realminarization mechanisms is their effect on the average turbulent velocity profile.

In order to test a possible connection between the initial flat velocity profile and the subsequent turbulence collapse, we carried out further computer simulations where this time a forcing term was added to the full Navier Stokes equations. The force was formulated such that it decelerates the flow in the central part of the pipe cross section while it accelerates the flow in the near wall region. The mass flux through the pipe and hence Re remain unaffected (see Methods equ.3 and Extended Data Fig. 5). Unlike in the experiments where the disturbance is applied locally and persists in time, here the forcing is applied globally. As shown in Fig. 2b, upon turning on the forcing with sufficient amplitude the initially fully turbulent flow completely relaminarizes. Hence a profile modification alone suffices to destabilize turbulence. Interestingly, the energy required for the forcing is smaller than the energy gained due to drag reduction (even for intermediate forcing amplitudes Extended Data Fig.7). In this case we therefore obtain a net energy saving already in the presence of the forcing (in experiments the saving is achieved downstream of the perturbation location). After removal of the forcing (see Extended Data Fig.11) turbulence fluctuation levels continue to drop exponentially and the flow remains laminar for all times. This effect has been tested for fully turbulent flow for Reynolds numbers between 3000 and 100 000 and in all cases a sufficiently strong force was found to lead to a collapse of turbulence resulting in drag decrease and hence energy saving of up to 95%.

We next investigate whether a profile modification on its own also relaminarizes turbulence in experiments. While body forces like that used in the simulations are not available (at least not for ordinary, non-conducting fluids), profiles can nevertheless be flattened by a local change in the boundary conditions. For this purpose one pipe segment is replaced by a pipe of slightly (4%) larger diameter which is pushed over the ends of the original pipe and can be impulsively moved with respect to the rest of the pipe (see Methods and Extended Data Fig. 5). The pipe segment is then impulsively accelerated in the streamwise direction and abruptly stopped, the peak velocity of the 300D long movable pipe segment is equal or larger (up to 3 times) than the bulk flow speed in the pipe. The impulsive acceleration of the near wall fluid leads to a flattened velocity profile (red profile in Fig.1c). Despite the fact that overall the fluid is accelerated and additional shear is introduced, after the wall motion is stopped (abruptly, over the course of 0.2 s) turbulence also in this case decays (see Fig.2c and supplementary movie 2). If on the other hand the wall acceleration is reduced, with wall velocities lower than 0.8U, turbulence survives. The impulsive wall motion is found to relaminarize turbulence very efficiently up to the highest Reynolds number (Re=40 000) that could be tested in the experiment (here the wall was moved at the bulk flow speed).

In turbulent wall-bounded shear flows, energy has to be transferred continuously from the mean shear into eddying motion, and a key factor here is the interplay between streamwise vortices (i.e. vortices aligned with the mean flow direction) and streaks. The latter are essentially dents in the flow profile that have either markedly higher or lower velocities than their surroundings. Streamwise vortices "lift up" low velocity fluid from the wall and transport it towards the centre (see Extended Data Fig. 8). The low velocity streaks created in the process give rise to (nonlinear) instabilities and the creation of further vortices. Key to the efficiency of this "lift-up mechanism" is that weak vortices suffice to create large amplitude streaks. This amplification process is rooted in the non-normality (*4*) of the linear Navier Stokes operator and its magnitude is measured by the so called transient growth (TG) (see also Extended Data Fig. 8 and 9).

Computing TG for the forced flow profiles in the DNS, we indeed observe that TG monotonically decreases with forcing amplitude (see Extended Data Fig.10) and it assumes its minimum value directly before turbulence collapses. Generally, the flatter the velocity profile the more the streak vortex interaction is suppressed, and in the limiting case of a uniformly flat profile the lift up mechanism breaks down entirely.

Revisiting the experiments, the velocity profiles of all the disturbed flows considered exhibit a substantially reduced transient growth (Fig. 1c). For the streamwise injection, amplitudes relaminarizing the flow also show the minimum amplification (Fig. 2d) while at lower and higher injection rates where turbulence survives the amplification factors are higher and above the threshold found in the simulations. Similarly for the moving wall at sufficiently large wall acceleration where relaminarization is achieved, the lift up efficiency is reduced below threshold, while at lower wall speeds it remains above.

Some parallels between the present study and injection and suction control in channels and boundary layers (*18–20*) can be drawn. While for boundary layers during the injection phase the drag downstream increases, during the suction it decreases. Suction applied to a laminar Blasius boundary layer leads to a reduction of the boundary layer thickness and this is well known to delay transition and push the transition location downstream.

The drag reduction achieved for the different methods used to destabilize turbulence is summarized in Fig.3. In each case the friction value before the profile modification corresponds to the characteristic Blasius law for turbulence (upper line) and after the disturbance it drops directly to the laminar Hagen-Poiseuille law. Hence the maximum drag reduction feasible in practice is reached (Fig.3b), and at the highest Reynolds numbers studied a 95% reduction is obtained. Although the numerical and experimental relaminarization methods affect the flow in different ways, the common feature is that the velocity profile is flattened.

In summary we have shown that fully turbulent flow can be destabilized by appropriate perturbations. As a consequence the entire flow becomes laminar, drastically reducing friction losses. The commonly used criteria for flows to be turbulent, such as a threshold in the shear rate and the Reynolds number (see (*23, 24*) for other criteria) do not explain the collapse of turbulence observed here. Key to relaminarization is a rearrangement of the mean turbulent profile that inhibits the vortex streak interaction. As shown, this can be achieved in a variety of ways, offering a straightforward target for practical applications where potentially pumping and propulsion costs can be reduced by 95% or more. The future challenge is to develop and optimize methods that lead to the desired profile modifications in high Reynolds number turbulent flows.

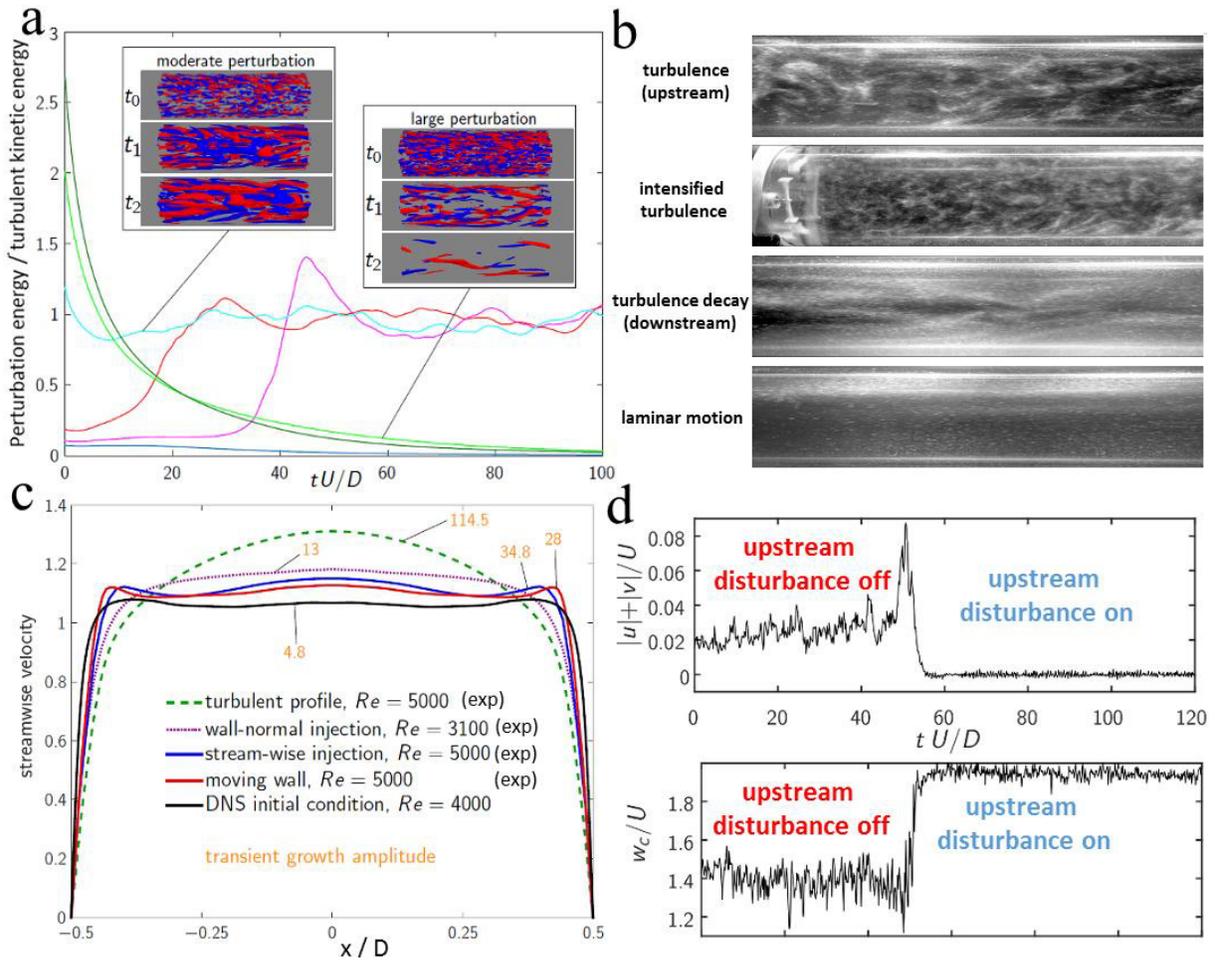

**Fig. 1. Perturbing turbulence: a** direct numerical simulations of pipe flow starting from turbulent initial conditions (taken from a run at Re=10 000), rescaled by a constant factor f and added to the laminar base flow at Re=4000, which was then integrated forward in time (at Re=4000). For small initial energies perturbations die out (dark blue curve). For sufficiently large energies (f≈1) transition to turbulence occurs (red, purple, cyan). For even larger energies (f>2.5) however the initially turbulent flow is destabilized and collapses after a short time (light and dark green curve). The 6 stream-wise vorticity isosurface figures show $\omega_z$ = +/− (red/blue) 7.2, 2.0, 1.6 U/D respectively at snapshot times $t_0$ = 0, $t_1$ = 5 and $t_2$ = 10 (D/U). **b** Fully turbulent flow (top panel) at Re=3100 is perturbed by vigorously stirring the fluid with two rotors. The more strongly turbulent flow (panel 2) eventually relaminarizes as it proceeds downstream (panel 3 and 4). **c** temporally and azimuthally averaged velocity fields of modified / perturbed flow fields in simulations and experiments. **d** Relaminarization of fully turbulent flow in experiments at Re=3100. The flow is perturbed by injecting 25 jets of fluid radially through the pipe wall. When actuated the fluctuation levels in the flow drop (top panel) and the centre line velocity switches from the turbulent level to the laminar value (2U), bottom panel.

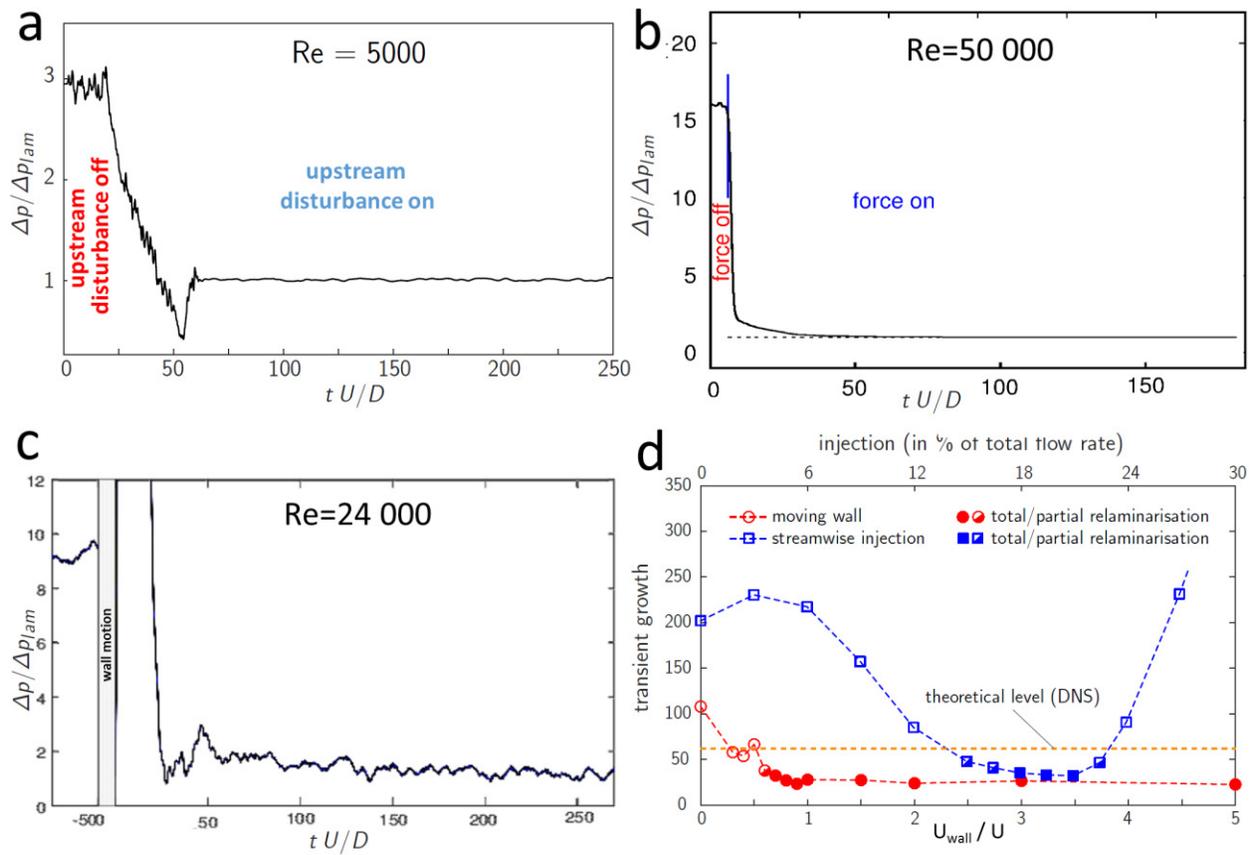

**Fig. 2.** Laminarization mechanisms. **a** After the streamwise near wall injection is actuated the pressure drop reduces to its laminar value. **b** A body force term is added in the numerical simulations which leads to an on average flatter flow profile (the fluid close to the wall is accelerated while it is decelerated in the near wall region). Disturbing the flow profile in this manner leads to a collapse of turbulence, here shown for Re=50000 where consequently friction losses drop by a factor of 10. **c** In the experiment the near wall fluid is accelerated via a sliding pipe segment, which is impulsively moved in the axial direction. Directly after the pipe segment is stopped the flow has a much flatter velocity profile. Subsequently turbulence collapses and the frictional drag drops to the laminar value. **d** Transient growth measures the efficiency of the lift up mechanism, i.e. how perturbations in the form of streamwise vortices are amplified while growing into streaks (deviations of the streamwise velocity component). All disturbance schemes used lead to a reduction in transient growth. The threshold value below which relaminarization occurs in the numerical (control via body force) is indicated by the orange line. For comparison the experimental flow disturbance mechanisms are shown in blue (streamwise injection) and red (moving wall). In agreement with the numerical prediction all disturbance amplitudes that lead to a collapse of turbulence (solid symbols) fall below the threshold value found in the simulations.

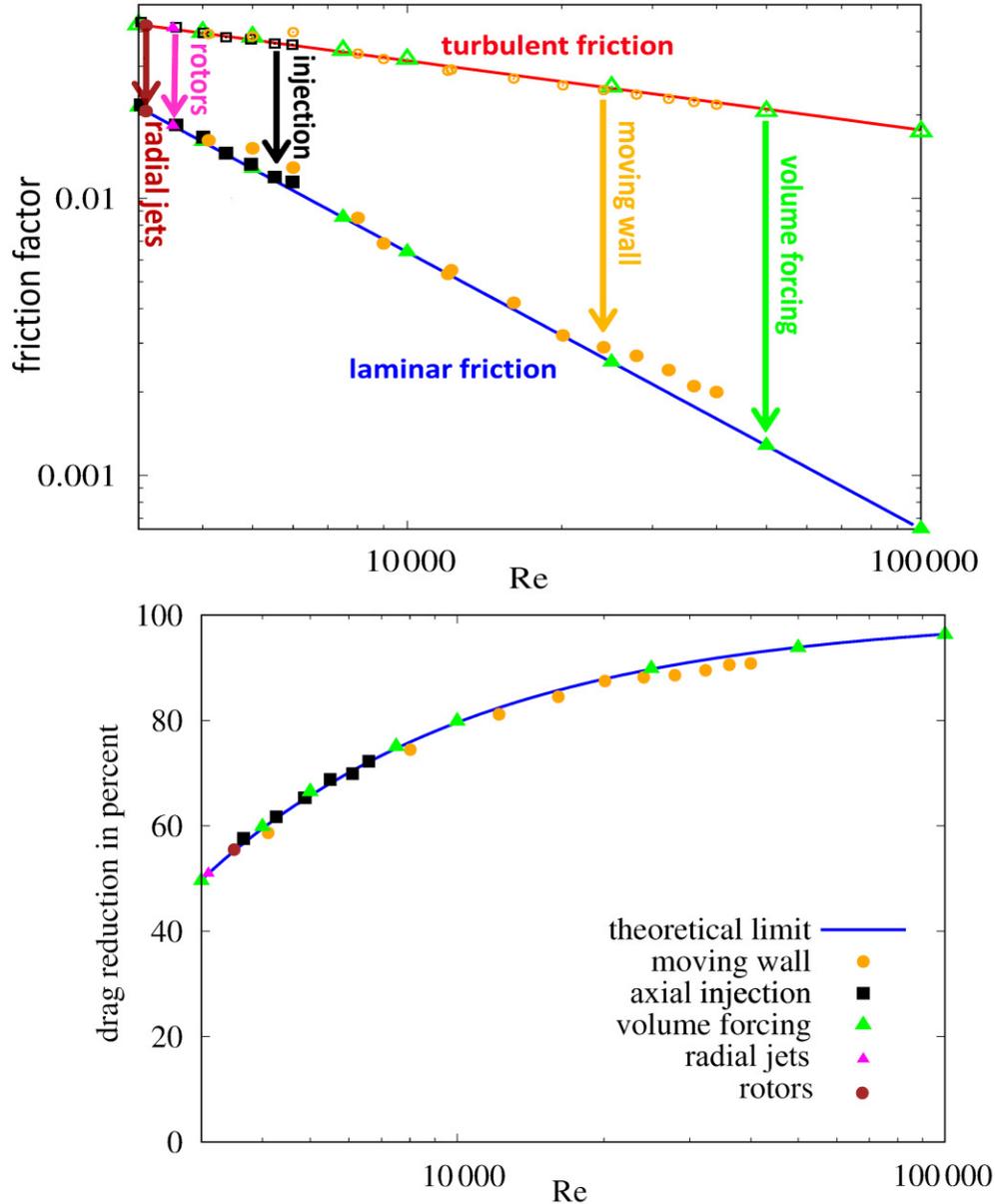

**Fig. 3.** Drag reduction. **a** Friction factor, f, as a function of Re. Initially all flows are fully turbulent and friction factors follow the Blasius-Prandtl scaling (f=0.316 Re$^{-0.25}$, red line). When the control is turned on flows relaminarize and the friction factors drop to the corresponding laminar values (Hagen- Poiseuille law in blue, f=64/Re). The rotors, radial jet injection, axial injection and moving wall controls are carried out in laboratory experiments while the volume force cases are from direct numerical simulations of the Navier-Stokes equations. For all cases the Reynolds number is held constant throughout the experiment. **b** Drag reduction as a function of Re. For the injection perturbation a maximum drag reduction of ~70% was reached whereas for the moving wall and volume forcing 90 and 95 % were achieved respectively. All data points reach the drag reduction limit set by relaminarization except for the Re>30000 in experiments where values are slightly above. Although these flows are laminar the profile shape is still developing and has not quite reached the Hagen Poiseuille profile yet (the development length required to reach a fully parabolic profile increases linearly with Re).

# 1 Methods

## 1.1 Experimental setup for the rotors

The facility consists of a glass-pipe (PMMA) with inner diameter $D = 54 \pm 0.2$ mm and a total length of 12 m ($222\,D$) made of 2 m sections (see Extended Data Fig. 1). The flow in the pipe is gravity driven and the working fluid is water which enters the pipe from a reservoir. The flow rate and hence the Reynolds number (Re $= UD/\nu$ where $U$ is the mean velocity, $D$ the diameter of the tube and $\nu$ the kinematic viscosity of the fluid) can be adjusted by means of a valve in the supply pipe. The temperature of the water is continuously monitored at the pipe exit. The flow rate is measured with an electromagnetic flowmeter (ProcessMaster FXE4000, ABB). The accuracy of Re is within $\pm 1\%$. To ensure fully turbulent flow the flow is perturbed by a small static obstacle (a 1 mm thick, 20 mm long needle located $10\,D$ after the inlet). 2 m downstream from the inlet the turbulent flow is perturbed by four small rotors which are mounted on a support structure within the pipe as indicated in Extended Data Fig. 1. The wiring of the motors is incorporated in the support structure of the motors. The rotors are small rectangular bars with even smaller rectangular bars at the tips. Their only purpose is to induce perturbations to the flow but no propelling motion or thrust. The rotors are turned at a rate of 7 rotations per second.

For the purpose of visual observations and video recordings of the flow field the flow is seeded with neutrally buoyant anisotropic particles (Matisse & Gorman 1984). The 3 locations where the supplementary movie (insert link) was recorded are indicated in Extended Data Fig. 1.

## 1.2 Experimental setup for the wall-normal jet injection and the streamwise injection through an annular gap

The facility consists of a glass-pipe with inner diameter $D = 30 \pm 0.01$ mm and a total length of 12 m ($400\,D$) made of 1 meter sections (see Extended Data Fig. 2). The flow in the pipe is gravity driven and the working fluid is water which enters the pipe from a reservoir. The flow rate and hence the Reynolds number (Re $= UD/\nu$ where $U$ is the mean velocity, $D$ the diameter of the tube and $\nu$ the kinematic viscosity of the fluid) can be adjusted by means of a valve in the supply pipe. The temperature of the water is continuously monitored at the pipe exit. The flow rate is measured with an electromagnetic flowmeter (ProcessMaster FXE4000, ABB). The accuracy of Re is within $\pm 1\%$. To ensure fully turbulent flow the flow is perturbed by a small static obstacle (a 1 mm thick, 20 mm long needle located $10\,D$ after the inlet). 2 m downstream from the inlet the turbulent flow can be perturbed in a controlled way by 2 different devices which are mounted within the pipe (see Extended Data Fig. 3 and Extended Data Fig. 4).

The velocity field is measured $\sim 330\,D$ downstream from the disturbance (control) at the position of the lightsheet. The measurement plane is perpendicular to the streamwise flow direction (pipe $z$-axis). All three velocity components within the plane are recorded using a high-speed stereo PIV system (Lavision GmbH) consisting of a laser and two Phantom V10 high-speed cameras with a full resolution of $2400 \times 1900\,\mathrm{px}$. The resulting spatial resolution is 77 vectors per $D$. The data rate is 100 Hertz. Hollow glass spheres (mean diameter $13\,\mu\mathrm{m}$, $\rho = 1.1\,\mathrm{g/cm}^3$) are used as seeding particles. Around the measurement plane the pipe is encased by a water filled prism such that the optical axes of the cameras are perpendicular to the air–water interface to reduce refraction and distortion of the images.

Downstream of the perturbation the pressure drop $\Delta p$ is measured between two pressure tabs with a differential pressure sensor (DP 45, Validyne, full range of 220 Pa with an accuracy of $\pm 0.5\%$) separated by $39.5\,D$ in the axial direction. As the difference in pressure drop between laminar and turbulent flows is very distinct even at moderate Reynolds numbers the signal is utilized to observe whether the flow is laminar or turbulent.

## 1.3 Experimental setup for the moving pipe

A movable perspex-pipe with inner diameter $D = 26 \pm 0.1\,\mathrm{mm}$ and a total length of 12 m ($461\,D$) is fitted to very thin-walled stainless steel pipes (MicroGroup) with outer diameter $d_{st,o} = 25.4 \pm 0.13\,\mathrm{mm}$ and a wall thickness of $0.4 \pm 0.04\,\mathrm{mm}$ such that the perspex pipe overlaps the steel pipes at the upstream and downstream end (see Extended Data Fig. 5). The steel pipes are stationary (mounted on fixed bearings). With respect to the perspex pipe they act as support and slide bearing, allowing the perspex pipe to be moved back and forth in the axial direction. To prevent sagging the perspex pipe is supported by 6 additional bushings (polymer sleeve bearings, Igus). To avoid leakage a radial shaft seal is mounted at both ends of the perspex pipe.

The length of the control section between stationary upstream and downstream stainless steel pipe, i.e. the actual length where the wall of the perspex pipe is in contact with the fluid and can be moved relative to the mean flow by moving the perspex pipe, is $L_{control} = 385\,D$.

The perspex pipe is connected to a linear actuator (toothed belt axis with roller guide driven by a servomotor, ELGA-TB-RF-70-1500-100H-P0, Festo; not shown in the figure). The linear actuator can move the perspex pipe for an adjustable distance (traverse path) $s \leq s_{max} = 1.5\,\text{m}$ at an adjustable velocity $U_{pipe} \leq U_{pipe,max} = 5.5\,\text{m/s}$. The maximum acceleration is $a = 50\,\text{m/s}^2$. The resulting wall velocity of the perspex pipe $u_{wall}$ is specified as a ratio to the mean flow velocity $U$, such that $u_{wall} = U_{pipe}/U$.

The flow rate and hence the Reynolds number (Re $= UD/\nu$ where $U$ is the mean velocity, $D$ the diameter of the tube and $\nu$ the kinematic viscosity of the fluid) can be adjusted by means of a valve in the supply pipe. The temperature of the water is continuously monitored at the pipe exit. The flow rate is measured with an electromagnetic flowmeter (ProcessMaster FXE4000, ABB). The accuracy of Re is within $\pm 1\%$. The flow is always turbulent when entering the control area. The velocity field is measured $\sim 50\,D$ upstream from the downstream steel pipe. The measurement plane is parallel to the streamwise flow direction (pipe $z$–axis) and located in the centerline of the pipe. The two velocity components within a plane of $\sim 3.5\,D$ length are measured using a high-speed 2D-PIV system (LaVision) with a full resolution of $2400 \times 1900\,\text{px}$. The resulting spatial resolution is 56 vectors per D. The data rate is 100 Hertz. Hollow glass spheres with a mean diameter of $13\,\mu\text{m}$ are used for seeding. Around the measurement plane the pipe is encased by a small rectangular perspex-box ($50 \times 50 \times 350\,\text{mm}$) filled with water such that the optical axis of the camera is perpendicular to the air–water interface to reduce refraction and distortion of the images.

A differential pressure sensor (DP 45, Validyne, full range of 550 Pa with an accuracy of $\pm 0.5\%$) is mounted onto the movable perspex pipe. Here the pressure drop $\Delta p$ in the perspex pipe is measured between two pressure taps (axial spacing 260 mm).

## 1.4 Energetic considerations

### 1.4.1 Wall-normal jets

We assume a fully-developed turbulent flow of density $\rho$ at $Re_b = 3100$, with mean velocity $U$ and Darcy friction factor (Pope 2001) $f_T = 0.0439$. The pipe diameter is $D = 30$ mm, whereas the injection hole diameter is $D_J = 0.5$ mm. The injection ratio is $\beta = NQ_J/Q = 0.015$, where $N = 25$ and $Q$ and $Q_J$ are the main and single injection flowrates. We start by expressing the kinetic energy per unit time pumped by the jets into the flow as a function of $Q$ and $U$. We have

$$N\Pi_J = N\frac{1}{2}\rho U_J^2 Q_J = \frac{1}{2}\rho U_J^2 \beta Q = \frac{1}{2}\rho U^2 Q \frac{\beta^3}{N^2}\left(\frac{D}{D_J}\right)^4. \quad (1)$$

Next, in order to compute the power associated to the controlled flow, we subdivide our domain in a control and remainder section. The control section has a length of $L_C = 25\,D$ and is characterized by a constant friction factor $f_C \approx 0.05$, estimated from the the velocity profile measured at the end of the jets section. The remainder length is $L_R = 330\,D$ with an average friction factor $f_R = 0.022$. The value is slightly larger than the laminar one as the relaminarizing flow takes about $140\,D$ to reach the a fully-developed state. We proceed by evaluating all the sources of power consumption. We have

$$\Pi_C = \frac{1}{2}\rho U^2 Q \frac{L_C}{D} f_C, \tag{2}$$

$$\Pi_R = \frac{1}{2}\rho U^2 Q \frac{L_R}{D} f_R, \tag{3}$$

$$\Pi_T = \frac{1}{2}\rho U^2 Q \frac{L_C + L_R}{D} f_T, \tag{4}$$

where $\Pi_C, \Pi_R, \Pi_T$ are the power dissipated in the control section, in the pipe remainder and in the whole pipe without control, respectively. The power difference with respect to the turbulent reference is

$$\Delta\Pi = \Pi_T - (\Pi_C + \Pi_R + N\Pi_J), \tag{5}$$

which implies a net saving if $\Delta\Pi > 0$. To quantify the efficiency of the control method we follow the same approach as in Kasagi et al. (2009), Quadrio (2011). We compute the relative net power saving

$$S = \frac{\Delta\Pi}{\Pi_T} = 45\%,$$

and the control cost normalized by the power gain,

$$\frac{1}{G} = \frac{N\Pi_J}{\Pi_T - (\Pi_C + \Pi_R)} = 1\%,$$

which also represents the minimum actuator efficiency required to have $S > 0$. As a final remark, it is possible to estimate the minimum remainder length that ensures a positive net gain. By repeating the estimate of $S$ for smaller values of the remainder length we find

$$L_R|_{S=0} \approx 25D.$$

### 1.4.2 Stream-wise injection

Similarly to the normal jets case, we estimate the efficiency of the stream-wise injection device. To this end, we assume a fully-developed turbulent flow of density $\rho$ at $Re_b = 5000$, with mean velocity $U$ and Darcy friction factor $f_T = 0.0377$. The ratio between the area from which the fluid is injected and the total area is $\beta_A = 0.13$, while the injection ratio is $\beta_Q = Q_J/Q = 0.18$. The kinetic energy per unit time introduced by the annular jet into the flow can be estimated as

$$\Pi_I = \frac{1}{2}\rho U_I^2 Q_I = \frac{1}{2}\rho U^2 Q \frac{\beta_Q^3}{\beta_A^2}. \tag{6}$$

An additional pressure loss due to the area contraction inside the injection device is modeled as

$$\Pi_L = \frac{1}{2}\rho U^2 Q K, \tag{7}$$

where $K = 0.5$ is chosen conservatively. The remainder section is identical to the wall-normal jets case but with friction factor $f_R = 0.0142$. The power difference with respect to the turbulent reference is

$$\Delta\Pi = \Pi_T - (\Pi_R + \Pi_L + \Pi_I). \tag{8}$$

Consequently, the relative net power saving and control cost normalized by the power gain are respectively

$$S = \frac{\Delta\Pi}{\Pi_T} = 55\%,$$

and

$$\frac{1}{G} = \frac{\Pi_I}{\Pi_T - (\Pi_C + \Pi_R)} = 5\%.$$

Finally, the minimum remainder length that allows a positive net power saving is estimated to be

$$L_R|_{S=0} \approx 47D,$$

## 1.5 Numerical method

We solve the incompressible Navier-Stokes equations

$$\frac{\partial \boldsymbol{u}}{\partial t} + \boldsymbol{u} \cdot \boldsymbol{\nabla} \boldsymbol{u} = -\boldsymbol{\nabla} p + \frac{1}{Re}\Delta \boldsymbol{u}, \quad \boldsymbol{\nabla} \cdot \boldsymbol{u} = 0 \tag{9}$$

in a straight circular pipe in cylindrical coordinates $(r, \theta, z)$, $r$, $\theta$ and $z$ being the radial, azimuthal and axial coordinate respectively. Throughout this study, the flow is driven by a constant mass flux. In Eqs. 9 velocity is normalized by the mean velocity $U$ and length by pipe diameter $D$. A Fourier-Fourier-finite difference code is used for the integration of the governing equations, with periodic boundary condition in the axial and azimuthal directions. In the radial direction a central finite difference scheme with a 9-point stencil is adopted. In this formulation, velocity can be expressed as

$$\boldsymbol{u}(r, \theta, z, t) = \sum_{k=-K}^{K} \sum_{m=-M}^{M} \hat{\boldsymbol{u}}_{k,m}(r, t) e^{(i\alpha k z + i m \theta)} \tag{10}$$

where $\alpha k$ and $m$ give wave numbers of the modes in axial and azimuthal direction respectively, $2\pi/\alpha$ gives the pipe length $L_z$, and $\hat{\boldsymbol{u}}_{k,m}$ is the complex Fourier coefficient of mode $(k, m)$. The governing equations are integrated with a $2^{nd}$-order semi-implicit time-stepping scheme, for details see Willis & Kerswell (2009). The code has been verified and extensively used in many studies (e.g., Willis & Kerswell (2009), Avila et al. (2010), Barkley et al. (2015)).

| $Re = \frac{UD}{\nu}$ | $Re_\tau = \frac{u_\tau D}{\nu}$ | $L_z(D)$ | $L_z^+$ | $\Delta r^+_{min}$ | $\Delta r^+_{max}$ | $(\Delta\theta D/2)^+$ | $\Delta z^+$ |
|---|---|---|---|---|---|---|---|
| 4000 | 141 | 12.6 | 1776 | 0.023 | 1.3 | 3.5 | 3.5 |
| 5000 | 171 | 12.6 | 2155 | 0.028 | 1.6 | 4.2 | 4.2 |
| 10000 | 314 | 6.3 | 1972 | 0.016 | 3.4 | 3.4 | 6.5 |
| 25000 | 700 | 3.0 | 2100 | 0.016 | 2.4 | 4.3 | 4.1 |
| 50000 | 1285 | 1.5 | 1927 | 0.010 | 2.5 | 5.2 | 5.0 |
| 100000 | 2357 | 0.8 | 1886 | 0.008 | 3.1 | 4.8 | 5.0 |

Extended Data, Table 1: The Reynolds number $Re$, the friction Reynolds number $Re_\tau = \frac{u_\tau D}{\nu}$ ($u_\tau = \sqrt{\frac{\tau_w}{\rho}}$ is the friction velocity defined based on wall shear stress $\tau_w$ and density $\rho$), pipe length in diameter, pipe length in wall unit $L_z^+ = \frac{L_z u_\tau}{\nu} = \frac{L_z}{D} Re_\tau$, the smallest and maximum radial grid size $\Delta r^+_{min} = \frac{\Delta r_{min}}{D} Re_\tau$, $\Delta r^+_{max} = \frac{\Delta r_{max}}{D} Re_\tau$, azimuthal grid size at the pipe wall (maximum) $(\Delta\theta D/2)^+ = \frac{\Delta\theta D/2}{D} Re_\tau$ and axial grid size $\Delta z^+ = \frac{\Delta z}{D} Re_\tau$ in wall unit.

In Table 1, we list the Reynolds numbers, pipe lengths and resolutions we considered in our simulations. To avoid significant domain size effect, the pipe lengths are selected to contain a few low-speed streaks, whose streamwise length is typically around 500 wall units in our normalization, see Jimenez & Pinelli (1999). The pipe length was doubled for $Re = 4000$ and $5000$ to verify that the pipe lenghs here in the table are sufficient. The resolutions are set to be able to sufficiently resolve the near wall structures, see reference grid sizes shown in the Table 1 of Jimenez & Pinelli (1999). Note that there is a difference of a factor of 2 in length scales between our normalization and theirs (double ours to compare with theirs).

### 1.5.1 Forcing the flow

To implement our control technique, i.e. to deform the velocity profile, an external force term $\boldsymbol{F} = F(r)\hat{\boldsymbol{z}}$ is introduced to the Navier-Stokes equations. This force decelerates the flow near the pipe center and accelerates the flow near the pipe wall while keeping the mass flux unchanged. As a result the velocity profile is deformed on average to a more plug-like one compared to the parabola $\boldsymbol{U} = (2 - 8r^2)\hat{\boldsymbol{z}}$ in the unforced situation. This forcing technique in essence is the same as in Hof et al. (2010), however a different functional form was chosen for the forcing in order to control fully turbulent flow. In this study, the force is such that it generates a velocity profile in laminar flow given by

$$\boldsymbol{u}(\beta, r) = (2 - \beta)(1 - \frac{\cosh{(2cr)} - 1}{\cosh{(c)} - 1})\hat{\boldsymbol{z}} \tag{11}$$

where $\beta$ is the decrease of the centerline velocity compared to the parabola and will be taken as a measure of the forcing amplitude. $c$ is a parameter to assure the constant mass flux. The body force $\boldsymbol{F}$ is solved inversely given the target profile.

As an example, a force and the resulting velocity profile at $Re = 5000$ with the force parameter $\beta = 0.6$ is shown in Extended Data Fig. 6. As shown, the body force is negative, i.e., acting upstream near the pipe center, slowing down the flow, and is the other way around near the pipe wall.

Subsequently, this body force will be imposed globally on top of fully turbulent flows. At high Reynolds numbers (above 3000), we observed that turbulence indeed decays and the flow relaminarizes given sufficient force, as shown in Fig.2b in the main text and in Extended Data Fig. 11 at $Re = 50000$ where the flow is forced with the force parameter $\beta = 0.98$. At this $Re$, the friction drops by a factor of 16 after the flow relaminarizes and the velocity profile recovers towards the parabolic Hagen-Poiseuille profile. The same control was tested up to $Re = 1 \times 10^5$ in our simulations and relaminarization was also obtained given sufficiently strong force.

Our simulations also show that an energy saving is immediately achieved when the force is activated, even if flow stays turbulent under the forcing. To illustrate this point, forces with several different amplitudes are tested at $Re = 4000$ and the results are shown in Extended Data Fig. 7. The energy consumption is calculated as the the power of the driving pressure gradient and the controlling force per unit volume (power density) as

$$P_\mathrm{p} = \frac{\int_V (-\nabla p \cdot \boldsymbol{u}) dV}{V} \tag{12}$$

and

$$P_\mathrm{F} = \frac{\int_V (\boldsymbol{F} \cdot \boldsymbol{u}) dV}{V}. \tag{13}$$

The integration is performed over the whole computational domain $V$. Note that the driving pressure gradient is spatially invariant and the controlling force is only radially dependent, therefore, these two forces only do work on the mean flow. The energy consumption due to the enhanced skin friction (flattened velocity profile) under the forcing is accounted for by the energy consumption of the controlling force. Consequently the energy saving is defined as

$$S = \frac{P_{\text{p, unforced}} - P_{\text{p, forced}} - P_{\text{F}}}{P_{\text{p, unforced}}}. \tag{14}$$

The time series of the energy consumption of an example at $Re = 4000$ for the driving pressure gradient and for the active forcing are shown in Extended Data Fig. 7a. We can see that, upon activating the forcing at $t = 100$, the decrease in the driving pressure gradient (energy gain) outweighs the energy consumption of the forcing (energy loss), meaning a net energy saving, though turbulence remains at a lower level under the forcing. The energy saving also increases with a stronger forcing, as shown in Extended Data Fig. 7b.

### 1.5.2 The lift-up mechanism

To better understand why turbulence decays in the presence of a disturbance (/forcing) that flattens the velocity profile we consider how the profile shape influences the lift-up mechanism. This mechanism is a major energy growth mechanism in shear flows and is responsible for the transition to turbulence in linearly stable shear flows (see a recent review by Brandt (2014)). It has been shown that, for inviscid flow, streamwise invariant cross-flow disturbances, such as streamwise rolls, do not decay and thus continually convect the mean shear (i.e. $u_r \frac{\partial U_z(r)}{\partial r}$) and redistribute the streamwise momentum, generating strong low/high speed streamwise streaks. The strongly distorted velocity profile becomes susceptible to other instabilities which generate 3-D turbulent fluctuations via nonlinear interaction (e.g., see Hamilton et al. (1995)). Theoretical argument of Landahl (1980) showed that the disturbance kinetic energy grows at least linearly with time in inviscid flows. This energy growth will be limited by the viscosity in viscous flows, however, only at large times. On the other hand, the lift-up ($u_r \frac{\partial U_z(r)}{\partial r}$) directly enters the turbulence production term in the equations for the kinetic energy of turbulent fluctuations (see e.g. Song et al. (2017)).

Extended Data Fig. 8 illustrates the lift-up exhibited by a vortex pair imposed on the mean turbulent flow profile. The vortices redistribute the shear and lift up slow fluid (blue) from the wall and replace it by faster fluid (red) from the central part of the pipe. The initial perturbation (consisting of the vortex pair) is strongly amplified as it is transformed into streaks. To obtain a measure of this amplification mechanism we consider the linearized Navier-Stokes equations and perform a transient growth (TG) analysis (following the analysis of Butler & Farrell (1993) and the algorithm by Meseguer & Trefethen (2003)). As the forced mean turbulent velocity profile is linearly stable small perturbations to the linearized equations must eventually decay. However disturbances of the form shown in Extended Data Fig. 8a will experience significant growth for some transient period (during lift-up) before they eventually decay (Extended Data Fig. 9). For the case shown the initial disturbance energy is amplified by a factor 70, i.e. the eventual streaks have a 70 times larger energy than the initial vortices. We next probe how TG is affected for the forced profiles. As shown in Extended Data Fig. 10, as the forcing amplitude is increased TG continuously decreases, i.e. by flattening the velocity profile the vortex streak interaction becomes less efficient. At a forcing amplitude of about 0.60 turbulence decays and the flow relaminarizes. The same procedure has been applied to the experimental flows: starting from the measured averaged velocity profile and assuming that the profile is fixed under the influence of the disturbance we conducted a TG analysis around this modified profile. As shown in Figure 2d in the main text the TG of profiles the disturbed profiles is indeed reduced considerably, suggesting that vortices are less efficient in producing streaks. Hence the energy growth via the lift-up mechanism is clearly subdued. As also illustrated in Figure 2d the collapse of turbulence in the experiments happens at comparable TG values as the ones found in the simulations for identical Re.

### 1.5.3 Removing the force

As shown in Extended Data Fig. 11, turbulence keeps decaying while a sufficient forcing is applied (black line after the vertical blue line). Under the forcing, turbulence continually decays and will eventually disappear. Clearly, the active forcing is consuming energy, manifesting higher shear at the wall than the ideal Hagen-Poiseuille flow (see Extended Data Fig. 6b), and it is not optimal if the force is always kept on. Thanks to the subcriticality of the laminar pipe flow, only perturbations above certain finite amplitudes can trigger turbulence, the force can be switched off once the tubrulent velocity fluctuations decayed below the critical value. After that, turbulence cannot recover even if the force is switched off. Here in the figure we show that after a sufficient control time (about 170 $D/U$ in this simulation), turbulence indeed keeps decaying when the force is removed (see the red solid line after about $t = 180$). Eventually the kinetic energy enters an exponential decay regime, which is the signature of a linear process. We did not continue the simulation due to the very high computational cost at this high Reynolds number. However, turbulence is not expected to recover in this linear regime.

Extended Data Fig. 12 shows the velocity profile at some time instants. The black line is the averaged velocity profile of a normal unforced turbulence at Re=50000. When the force is turned on at $t = 7.5$, the velocity profile is quickly flattened into a plug-like one as the blue line shows. After the force is switched off at $t = 180$, the velocity profile starts to recover towards the parabolic profile of the Hagen-Poiseuille flow because turbulent fluctuations are nearly extinct. However, this recovery is a long asymptotic process and roughly takes hundreds of convective time units. The red line shows the velocity profile during this recovery process at $t = 340$.

## 2 Extended Data

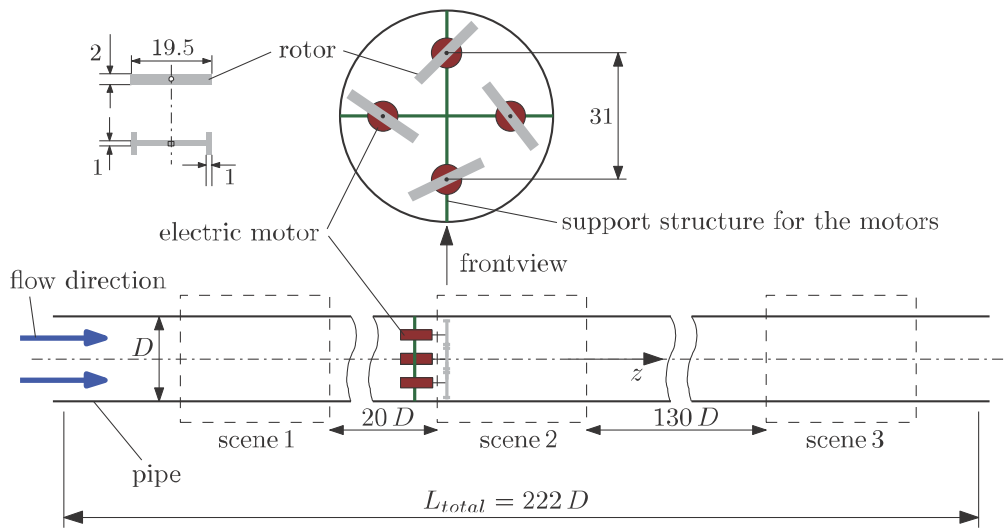

Extended Data, Figure 1: Sketch of the test facility used to perturb the flow with four rotors driven by four electric motors. The diameter of the PMMA-glass pipe is $D = 54\,\text{mm}$. The flow direction is from left to right. Drawing not to scale. All dimensions in mm. The dashed rectangles indicate the locations of the scenes in the movie.

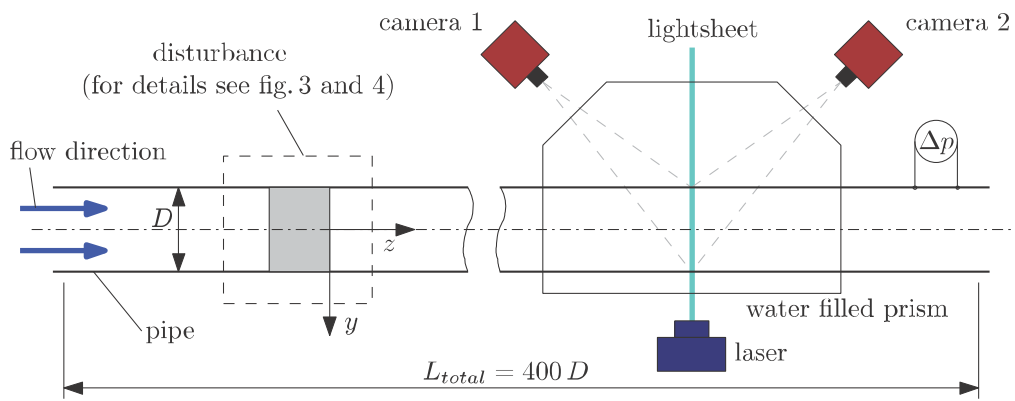

Extended Data, Figure 2: Sketch of the test facility used to disturb the flow with the wall-normal jet injection and the streamwise injection through an annular gap, see Extended Data Fig. 3 and Extended Data Fig. 4 for details. The diameter of the glass pipe is $D = 30\,\mathrm{mm}$. The flow direction is from left to right. Drawing not to scale.

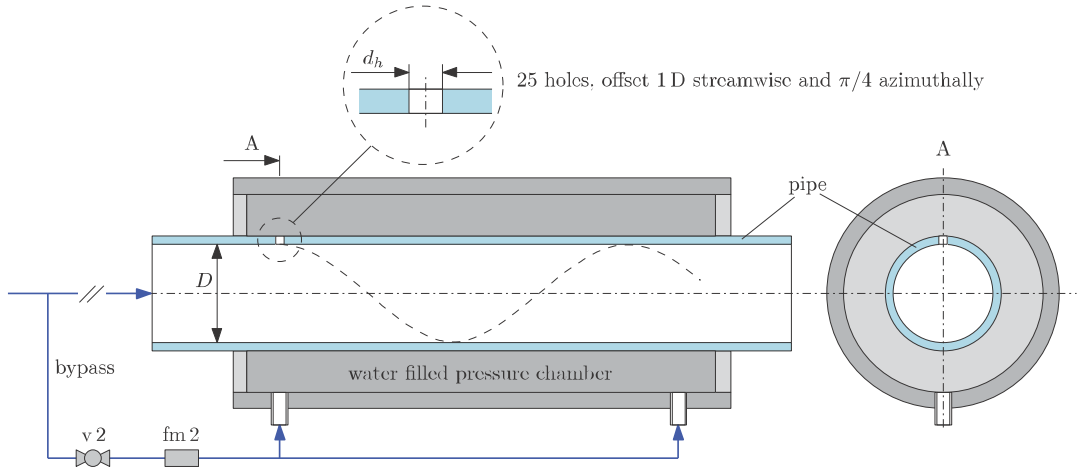

Extended Data, Figure 3: Device to inject wall-normal jets: Sketch of the device which allows to inject fluid into the main pipe through 25 small holes with diameter $d_h = 0.5$ mm. The holes are offset by $1\,\mathrm{D}$ in the streamwise and $\pi/4$ in the azimuthal direction. The section where the pipe is perforated is made of plexiglas and encased by a water-filled pressure chamber. To ensure uniform injection all the holes have been machined with a tolerance of 0.01 mm and are subjected to the same pressure, the latter guaranteed by the large volume of the water-filled encasing chamber. Fluid is taken from the main pipe via a bypass and then re-injected through the pressure chamber. The bypass is equipped with a valve (v 2) and a flowmeter (fm 2) to precisely adjust and measure the injected flow such that the flow rate and hence the velocity of the resulting jets can be precisely adjusted and measured. At $Re_b = 3100$ and for a by-pass ratio $\beta = 0.015$ the injection flow rate is $Q_J = 2.5$ ml/min per single hole, corresponding to an injection velocity $U_J = Q_J/(\pi d_h^2/4) = 0.2$ m/s. The device is mounted within two pipe segments. The flow direction is from left to right. Drawing not to scale.

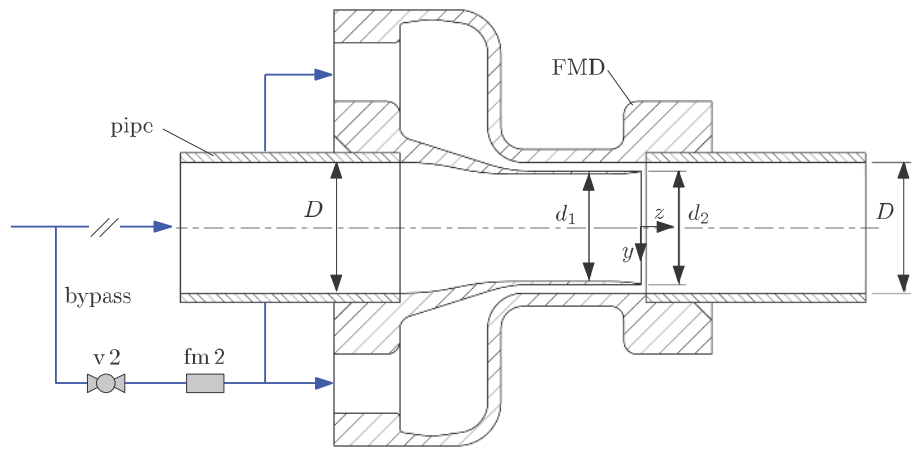

Extended Data, Figure 4: Device for streamwise injection through an annular gap: The device allows to inject fluid into the main pipe through a small annular concentric gap of 1 mm close to the pipe wall. The main pipe is slightly narrowed in a short range just upstream the injection point ($d_1 = 26.6$ mm, $d_2 = 28$ mm, open gap area $A_1 = 91.1$ mm$^2$). At a small backward facing step ($z = 0$, the plane of confluence) the fluid coming from the bypass is axially injected into the main pipe through an annular gap. The specified Reynolds number in the main pipe applies to the range upstream the bypass and downstream the confluence at $z = 0$. Fluid is taken from the main pipe via a bypass and then re-injected through a concentric gap close to the wall. The device is mounted within two pipe segments. The bypass is equipped with a valve (v 2) and a flowmeter (fm 2) to precisely adjust and measure the injected flow through the gap. The flow direction is from left to right. Drawing not to scale. Patent pending.

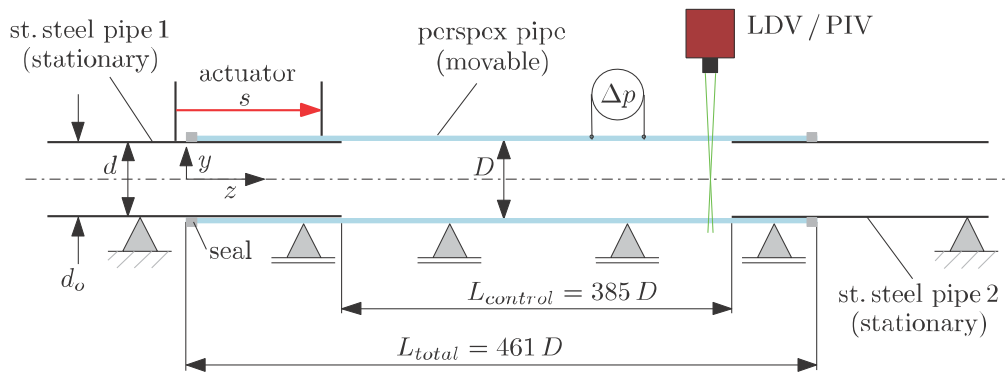

Extended Data, Figure 5: Sketch of the experimental setup with a moving pipe section. The flow direction is from left to right. A perspex pipe is slipped over two stationary, very thin walled stainless steel pipes such that the perspex pipe overlaps the steel pipes at the upstream and downstream end. The perspex pipe is movable in the axial direction for an adjustable distance $s$. Drawing not to scale.

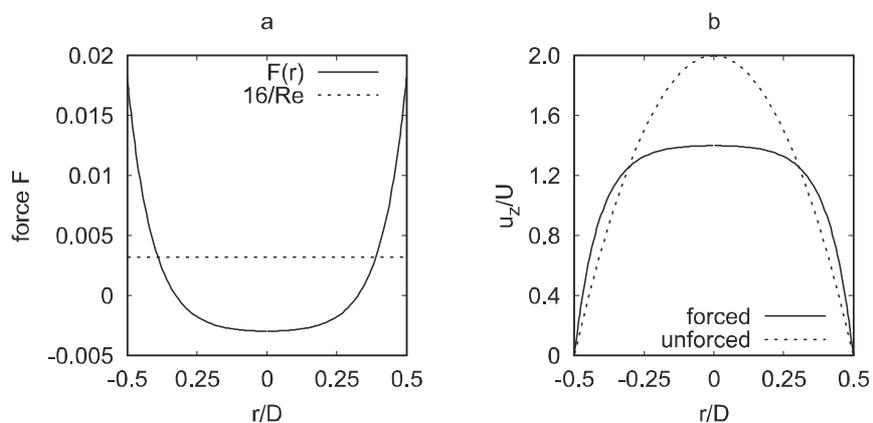

Extended Data, Figure 6: (a) The force $F(r)$ as a function of $r$ at $Re = 5000$ given the parameter $\beta = 0.6$ (solid line). For comparison, the pressure gradient of the unforced laminar flow $\partial p/\partial z = 16/Re$ is given by the dashed line. (b) The forced (solid line) and the unforced (dashed line) velocity profile of the basic laminar flow. Note that in the forced flow case both the $F(r)$ and pressure gradient $16/Re$ shown in (a) are at work.

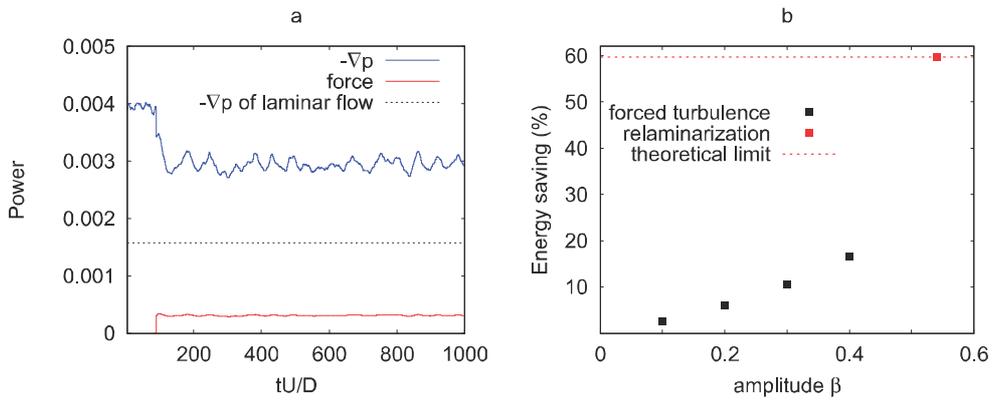

Extended Data, Figure 7: (a) The power signal of a forcing case at $Re = 4000$ with the parameter $\beta = 0.44$. The blue line is for the driving pressure gradient (see Extended Data Eqs. (12)) and the red for the active forcing (see Extended Data Eqs. (14)). The dashed line marks the power of the pressure gradient corresponding to laminar flow. Flow is initially fully turbulent and the force is turned on at $t = 100$. (b) Net energy saving as a function of the forcing amplitude $\beta$ at $Re = 4000$. At sufficient forcing amplitude, turbulence collapses and the energy saving approaches the theoretical limit of 60% at this Reynolds number (red symbol and dashed lines).

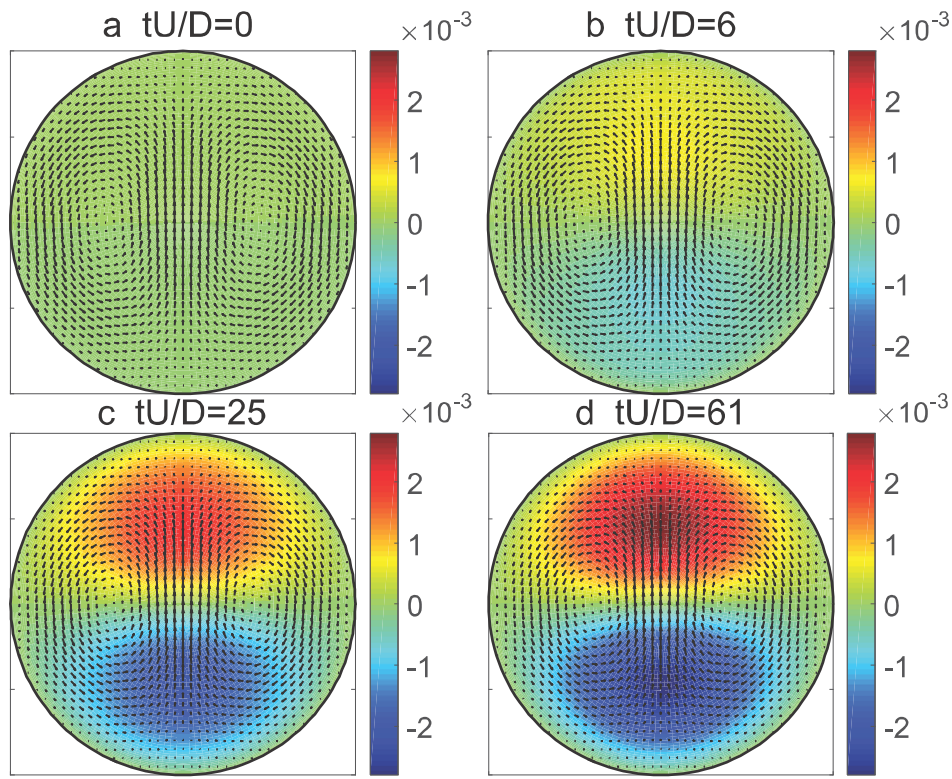

Extended Data, Figure 8: The generation of strong streaks by weak streamwise vortices due to the non-normality of the linearized flow. The calculation is at $Re = 5000$ with the parameter $\beta = 0.3$. The basic velocity profile is taken as the average velocity profile of the forced turbulent flow under this forcing parameter. A weak perturbation containing a pair of streamwise vortices (black arrows) is introduced to the flow and the growth of the streaks (colormap) are monitored (see the sequence of A-D). In the figure, the amplitude of the vortices (the maximum of the velocity component) is about $3 \times 10^{-4}$, which does not show significant change in all 4 panels however generates strong streaks.

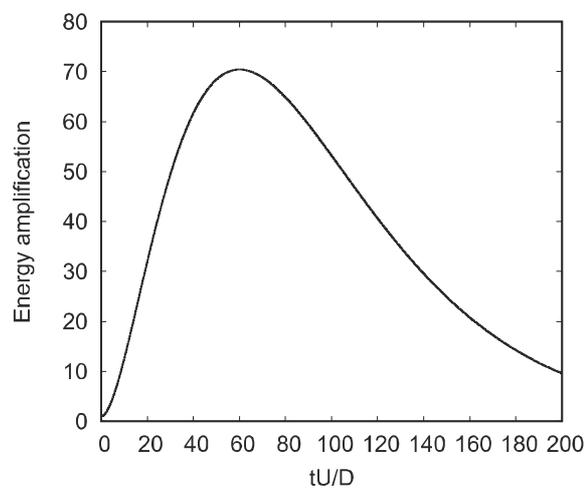

Extended Data, Figure 9: The transient amplification of energy of the flow (streaks) generated by the weak vortices as shown in Extended Data Fig. 8A. Large transient amplification can be reached before the disturbances eventually decay.

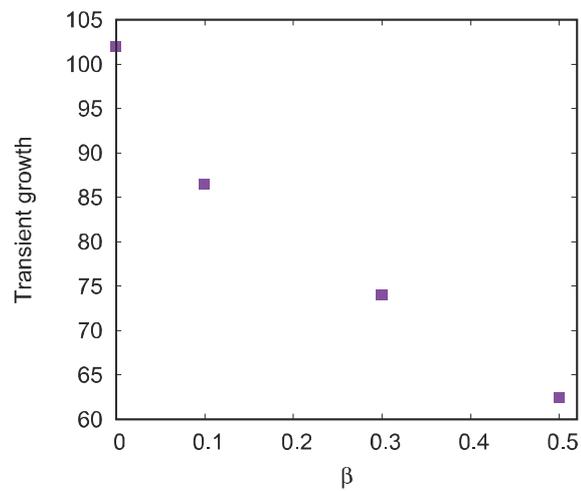

Extended Data, Figure 10: The transient growth of the forced turbulent mean velocity profile as the forcing parameter $\beta$ increases. Clearly, as $\beta$ increases, the transient growth decreases, i.e., the streaks generated by weak vortices become weaker. When $\beta$ surpasses about 0.5, turbulence tends to become localized and eventually relaminarizes as $\beta$ increases further.

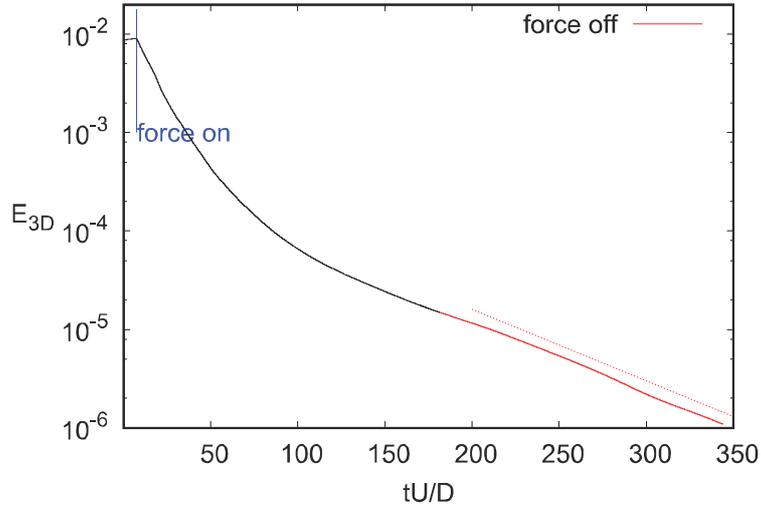

Extended Data, Figure 11: The kinetic energy of the 3-D turbulent fluctuations at Re=50000. Force $F(r)$ with $\beta = 0.98$ is turned on at $t = 7.5$ (marked by blue vertical line), and it is turned off at $t = 180$ and the kinetic energy after this point is shown as red line. The dotted red line indicates an exponential decay.

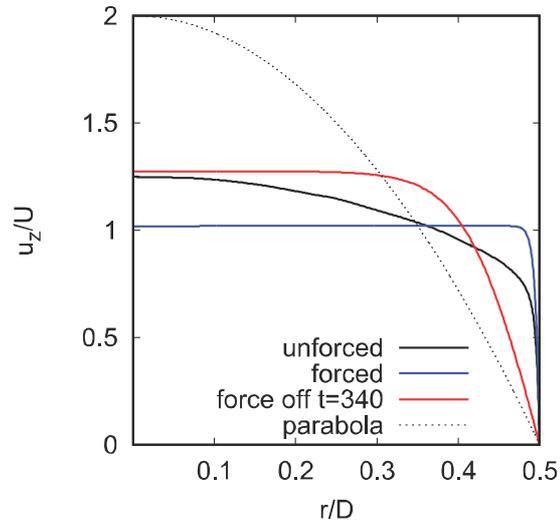

Extended Data, Figure 12: The velocity profile of the turbulence at Re=50000 as shown in Extended Data Fig 11 at a few time instants. The black one is the velocity profile of a normal turbulence, the blue one is for the forced velocity profile at $t = 100$. The red one is the velocity profile at $t = 340$ after the force was switched off at $t = 180$. The dashed black line is the parabolic profile of the Hagen-Poiseuille flow.

11